\documentclass[journal]{journal}
\usepackage{cite}
\usepackage{amsfonts}
\usepackage{booktabs} 
\usepackage[pdftex]{graphicx}

\newcommand{\mat}[3]{\ensuremath{\mathbf{#1}}_{#2\times#3}}
\newcommand{\mate}[3]{\ensuremath{\mathbf{#1}}_{#2#3}}

\begin{document}

\title{Supplier Recommendation in Online Procurement}
\author{Victor Coscrato and Derek Bridge\\ School of Computer Science \& Infomration Technology,\\ University College Cork, Ireland}

\maketitle
\thispagestyle{empty}

\begin{abstract}
Supply chain optimization is key to a healthy and profitable business.
Many companies use online procurement systems to agree contracts with suppliers.
It is vital that the most competitive suppliers
are invited to bid for such contracts. In this
work, we propose a recommender system to assist with
supplier discovery in road freight online procurement. 
Our system is able to provide personalized
supplier recommendations, taking into account customer needs and preferences.
This is a novel application of recommender systems, calling for design choices that fit the unique requirements of online procurement. Our preliminary results, using real-world data, are promising.
\end{abstract}


\section{Introduction} \label{sec:intro}

\IEEEPARstart{T}{he} world has been experiencing increasing
globalization. It is common to find companies that trade
their goods and services across several countries or
even continents. While expanding to new markets
brings benefits to businesses, maintaining good
supply chains has become an increasingly
difficult problem. In order to
keep factories and warehouses supplied with the raw materials and services that they
need, investments 
in logistic intelligence have been growing
\cite{Kherbash-Mocan-2015}.

A company that needs to purchase goods or services will sign contracts after an auction or after soliciting quotations. In both cases, several contending suppliers are invited to place
bids or to present quotations. After receiving bids or quotations, the purchaser
chooses one or more suppliers. This competition
helps the purchaser fulfil all their needs at the best rates and with the best quality guarantees.
In order to
have good options to choose between, it is vital the purchaser
invites bids or quotations from the best possible set of candidate suppliers.
In summary, a procurement process follows the
following steps: (i) requirements are listed, (ii) potential suppliers
are invited to bid or to present quotations,
(iii) bids and quotations are received, (iv) a supplier or a combination of several
suppliers that can fulfil the requirements are selected, and (v) contracts are awarded
to the chosen suppliers. We will use the word \textit{event} to refer to each instance of this process.

Increasingly, procurement events are run online. There are even companies that specialize in assisting other companies to run online procurement events. For example, the authors of this paper have worked with a company that provides an online platform which covers all the phases of a procurement process and which also offers a number of supply chain optimization tools. There is an opportunity, however, to expand this set of tools with a Supplier Recommender System. The recommender system will assist a purchaser in the vital task of selecting a good set of candidate suppliers for an event; some or all of the recommended suppliers can then be invited to submit bids or quotations. The supplier recommendations should be `personalized' to the event. 

In this paper, we propose a preliminary Supplier Recommender System. In particular, we use a Factorization Machine \cite{Rendle-2010}. These are suitable for this application since they are amongst the recommendation algorithms that can make use of both interaction data and meta-data. The interaction data is historical data of past events in which purchasers and suppliers participated; the meta-data describes the events ---for example, the locations involved, the type of goods involved, and so on. For reasons that we explain later, cold-start is a particular challenge in this domain.

In this paper, we report preliminary results that we obtain when training and testing the
system on real-world data that describes road freight events.
Road freight events are ones that can strongly benefit from a Supplier Recommender System since there are typically many possible suppliers (e.g.\ large road haulage companies, small local ``man-with-a van'' businesses, and so on) for this type of event. By contrast, there are typically fewer possible suppliers in the case of ocean freight.

Our contributions are:
\begin{itemize}
\item we show one possible way in which supplier recommendation can be represented as a recommendation problem;
\item we propose one recommendation algorithm (Factorization Machines) that can be used for the task; and
\item we evaluate on real-word data.
\end{itemize}
Our preliminary results show that there is great promise in tackling this problem in the way that we propose.

The paper proceeds as follows: Section \ref{sec:rw} discusses related work; Section \ref{sec:problem} shows how we have formulated supplier discovery as a recommendation problem; Section \ref{sec:alg} describes the recommender algorithm that we have adopted for this preliminary investigation;  Section \ref{sec:expts} describes our dataset, and the methodology and metrics that we use in a set of experiments; Section \ref{sec:results} discusses the results of the experiments; Section \ref{sec:disc} contextualizes the contribution through a discussion of some broader issues; and Section \ref{sec:concs} concludes the paper.

\section{Related Work} \label{sec:rw}

There is only a small amount of published research that is related to supplier recommendation. The problem is very domain-specific and so the existing literature
in the field is quite diverse. Jain et al.\ propose
the use of association rule mining to assist in the assessment of suppliers \cite{Jain-Et-Al-2007}. 
Nandeesh et al.\ also mine association rules but they include a second step that
evaluates the suppliers according to their price, the
delivery time, and the availability of the desired materials
or services \cite{Nandeesh-Et-Al-2015}. Lee et al.\ propose a two-step solution to the problem
of recommending sellers to customers in an open, online market: first a classification of the sellers
into trustworthy or not, followed by a content-based
recommendation algorithm to select from among the trustworthy sellers \cite{Lee-Et-Al-2013}.
Finally, Cui et al.\ conduct a study into the impact of AI on prices in procurement systems \cite{Cui-Et-Al-2021}. 

\section{Problem Formulation} \label{sec:problem}

Our goal is to develop a Supplier Recommender System (SRS) that can be used by a company that offers supply chain optimization services to its customers (referred to as purchasers). A purchaser creates an event, and the SRS recommends candidate suppliers. To achieve this objective, we use interaction data about previous events and meta-data about previous events and the new event. We explain these in more detail here.

In more traditional recommender systems, a ratings matrix $\mat{R}{U}{I}$ records user-item interaction data; for example, $\mate{R}{u}{i} = 1$ means that user $u$ watched movie $i$ or that user $u$ visited the web page for product $i$. $\mate{R}{u}{i} = \bot$ means that we have no record of any such interaction. It might be thought that in our SRS, purchasers play the part of users and suppliers play the part of items. But, in fact, events play the part that users play in traditional recommender systems: we are recommending suppliers to events. Hence, the interaction data takes the form of a participation matrix $\mat{P}{E}{S}$, where $E$ is the number of past events and $S$ is the number of suppliers. Now, $\mate{P}{e}{s} = 1$ means that supplier $s$ participated in event $e$. In contrast to traditional recommenders, this means we can interpret $\mate{P}{e}{s} = \bot$ as negative information: it means that $s$ did not participate in event $e$.

In our preliminary work, a supplier is considered to have participated in an event (i.e.\ $\mate{P}{e}{s} = 1$) if they were invited to bid or to give a quotation. This is a pragmatic decision based on the fact that this was the only data being kept at the moment by the company that we are working with. In future, we might hope to distinguish between suppliers who were invited to bid, those who did bid, those with whom contracts were signed, and those who fulfilled their contract satisfactorily. Hence, in future, matrix $\mathbf{P}$ might no longer be unary-valued.

In addition to the interaction data, we also have meta-data about past events, $\mat{X}{E}{M}$, where $M$ is the number of event features. The identity of the purchaser is considered to be one of the pieces of meta-data that describes an event and so the features include a one-hot encoding of the purchaser identity. Our meta-data also includes simple features such as the event timezone. Finally, the meta-data includes a short textual description of the event. Hence, our features include a bag-of-words vector representation of these textual descriptions. 

The supplier recommendation problem in online procurement
is especially difficult because recommendations
have to be provided for newly-created
events. In general, the system has no 
interaction data about the new event. It is the equivalent in a more traditional recommender systems of the cold-start user problem: how to recommend to users for whom we have no interaction data \cite{Gope-Jain-2017}. We will, however, have meta-data about the new event (e.g.\ who the purchaser is and a textual description). While this suggests that we should use a content-based recommender system that exploits the meta-data of the new event and the past events, we want a solution that uses the historical participation data too. For example, we may know that a particular supplier tends to participate in events with a particular purchaser or that a particular supplier tends to participate in events that have a particular description. So, ideally, we want an approach that makes use of both $\mathbf{P}$ and $\mathbf{X}$. We describe a recommendation model that can do this in the next section. 

\section{Recommender Algortihm} \label{sec:alg}

We use a Factorization Machine (FM) of degree $D=2$ \cite{Rendle-2010} as our recommendation
algorithm. FMs are among the models that are  especially appealing
for our problem because they can take advantage
of both the interaction data (in our case $\mathbf{P}$) and the event meta-descriptions ($\mathbf{X}$).
Furthermore, FMs have been shown to perform well in multiple
recommendation tasks \cite{Rendle-Et-Al-2011,Loni-Et-Al-2014}.

FMs can be thought of as a generalization of 
popular matrix factorization recommendation algorithms,
such as FunkSVD \cite{Funk-2006}. The concept
behind both these kinds of algorithms is the representation
of the entities (in our case, events and suppliers)
in a latent vector space. Furthermore, the
generalization brought by FM allows features
to be represented in this same latent space.

We have one training instance for each entry in $\mathbf{P}$, hence each training instance represents the fact that a certain supplier $s$ participated in a certain historical event $e$. The training instance comprises the one-hot encoded suppliers vector,
$\mathbf{1}_s$, and the event’s features $\mathbf{X}_e$. Recall that these features include the one-hot encoded purchaser vector $\mathbf{1}_p$, some simple features such as the event timezone, and the bag-of-words vector that describes the event. For convenience, we designate an arbitrary training instance as $\mathbf{x} = [\mathbf{1}_s \mathbf{X}_e]$, i.e.\ the concatenation of $\mathbf{1}_s$ and $\mathbf{X}_e$, with
dimension $n = S + M$. 
 
The predicted relevance score of an instance $\mathbf{x}$ has the form:
$$\hat{r}(\mathbf{x}) = \mathbf{w}_0 + \sum_{i=1}^n \mathbf{w}_i\mathbf{x}_i +
\sum_{i=1}^n\sum_{j=i+1}^n\langle \mathbf{v}_i,\mathbf{v}_j\rangle \mathbf{x}_i\mathbf{x}_j$$
The parameters are the intercept $\mathbf{w}_0 \in \mathbb{R}$, the linear coefficients $\mathbf{w} \in \mathbb{R}^n$, and a matrix $\mathbf{V} \in \mathbb{R}^{n\times D}$. $\langle \mathbf{v}_i,\mathbf{v}_j\rangle$ is the dot product of rows $\mathbf{v}_i, \mathbf{v}_j \in \mathbf{V}$, i.e.\ $\sum_{f=1}^D \mathbf{v}_{i,f}\mathbf{v}_{j,f}$, and uses a latent representation to model the two-way interactions between features $\mathbf{x}_i$ and $\mathbf{x}_j$. We refer to all the model parameters as $\Theta$.

We use Bayesian Personalized Ranking (BPR) \cite{Rendle-Et-Al-2011}
as our rank-aware loss function. The BPR loss function requires the training (positive) instances to be compared against sampled negative ones. A negative sample $\mathbf{x}^{-}$ is
composed of $[\mathbf{1}_s^* \mathbf{X}^e]$, where $\mathbf{1}_s^*$ one-hot encodes a
randomly-sampled supplier $s$ who did not participate in historical event $e$. 
Let $S_e$ denote the set of suppliers who participated in event
$e$. Then, the log-likelihood of the observed preferences 
is:
$$\sum_{e \in E}\sum_{\mathbf{x}^+ \in S_e}\sum_{\mathbf{x}^- \not\in S_e} \log \sigma(\hat{r}(\mathbf{x}^+) - \hat{r}(\mathbf{x}^-))$$
where $\sigma$ is the sigmoid function. Moreover, the
model parameters are assumed to have, a priori,
an independent multivariate Gaussian distribution,
that is $\Theta \sim N(0,\lambda I)$. The optimal $\Theta$ is chosen
by maximizing the log-likelihood of the posterior
distribution of $\Theta$ through stochastic gradient descent. In practice, for
each training instance $\mathbf{x}^+$, $l$ negative instances are
sampled, where $l$ is a hyper-parameter.

We take advantage of libFM \cite{Rendle-2012},
which is the original C++ implementation of the
FM algorithm, and a BPR extension of it \cite{Petroni-Et-Al-2015}.

\section{Experiments} \label{sec:expts}

We describe our dataset, methodology and metrics.

\subsection{Dataset}

As previously stated, our recommender system
is targeted at road freight events. This 
domain is characterized by the availability of many suppliers of considerable
diversity, ranging from
small trucking companies to major logistic service
providers. 
We use a real-world road freight event dataset. It has only a modest number of events ---a result of the fact that collecting data for recommendation purposes has not, up to now, been a focus of companies that operate in this space. However, there are many suppliers in the dataset, each participating in a small number of events.
We want to guarantee that every supplier will be represented both in the training
and test set, at least once. For this reason, we removed suppliers
who had participated in only one event.
The dataset also contained several events that
were created for testing purposes, e.g.\ used by a
customer to explore the user interface.
We removed these dummy events because often they do not
reflect real use cases.
Among the event features available, the recommender
system is currently using the event timezone, the auction type (either an e-auction
or a request for quotations), and a one-hot encoding of the purchaser company that created the event. Text descriptions of events were tokenized, lemmatized and represented as bag-of-word vectors using NLTK \cite{Bird-Et-Al-2009}.
Our final dataset consists of matrix $\mathbf{X}$ having 165 road freight
events with 495 event features; and matrix $\mathbf{P}$ in which there are records of 7023 cases of interaction between 1690 suppliers and the 165 events. Hence, the participation matrix has
a sparsity of 97.48\%.

\subsection{Method}

As explained earlier, the normal use case of this recommender will be to recommend suppliers who could be asked to participate in a brand new event. For
this reason, we want to evaluate our system in cold-start scenarios. Hence, we hold out entire
events, treating them as test instances and evaluating
them according to the metrics defined later.

As the dataset is quite small, we use 
nested cross-validation (CV). The
outer CV loop iterates through 8 equally-sized and
randomly-sampled folds of events, leaving one
evaluation fold out on each iteration. The inner loop is
a 5-fold CV used to tune the FM hyperparameters
(the latent dimension size $D$, the number of training iterations, the learning
rate, the regularization strength, and the number of
negative samples per positive $l$). The results are reported
as the average values across the outer CV
loop.

For comparisons, we use two other recommenders. The first is a non-personalized recommender that simply recommends the most popular suppliers. The second is an ablated version of the FM where training instances include only the supplier and purchaser identities $[\mathbf{1}_s, \mathbf{1}_p]$ and none of the other pieces of meta-data. This latter model is the closest we come to a standard matrix factorization model. It still recommends suppliers to events, but it has knowledge of the way that suppliers and purchasers participate in events. What it does not have is the other meta-data: these features are available only to our main FM model.

\subsection{Metrics}

We evaluate the top-$k$ recommendation of both the FM and the baselines for different values of $k$ using three metrics: precision,
recall and NDCG. In the interests of explicitness and reproducibility, and acknowledging the influence of \cite{Tamm-Et-Al-2021}, we will give the definitions that we have used for these metrics.

Let $\mathit{Train}$ and $\mathit{Test}$ be the set of training and test events, respectively. Let $S_e$ be the
set of suppliers who participated in $e \in \mathit{Test}$ --- in other words, this is the ground-truth. Then, let $\mathit{TopK}_e$ be the k-sized ordered
list of recommendations for event $e \in \mathit{Train}$. Then, the
precision and recall for test event $e$ are calculated as:
$$\mathit{Precision}@k(e) = \frac{|\mathit{TopK_e} \cap S_e|}{k}$$
$$\mathit{Recall}@k(e) = \frac{|\mathit{TopK}_e \cap S_e|}{|S_e|}$$

Discounted Cumulative Gain (DCG) \cite{Kanoulas-Aslam-2009} is a rank-sensitive metric, meaning
that it will reward correct recommendations
appearing earlier in the recommendation list. It is
calculated as follow:
$$\mathit{DCG}@k(e) = \sum_{i=1}^k\frac{I(\mathit{TopK}_e^{(i)} \in S_e)}{\log_2(i+1)}$$
where $\mathit{TopK}_e^{(i)}$  is the $i$-th recommended item in $\mathit{TopK}_e$ 
and $I(\cdot) \rightarrow \{0, 1\}$ is an indicator function. The Normalized
Discounted Cumulative Gain (NDCG) is a
corrected version of the DCG that bounds the metric
to the interval $[0,1]$ by dividing the DCG by its theoretical
highest value.

We compute the mean of these metrics, averaged across each $e \in \mathit{Test}$.

\section{Results} \label{sec:results}

\begin{figure}[t] 
\includegraphics[width=0.4\textwidth]{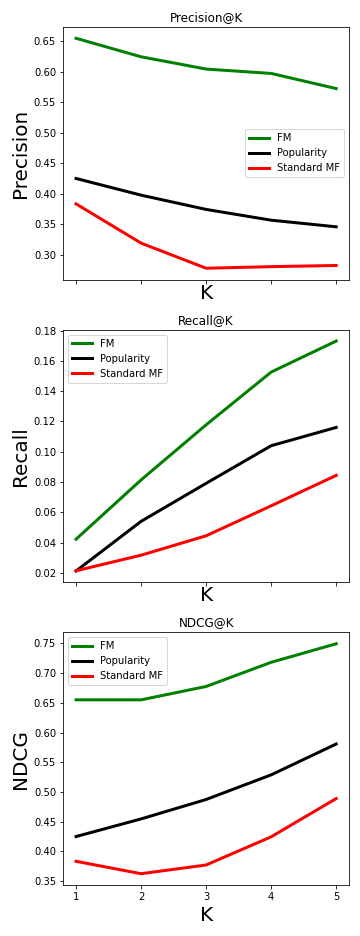}
\caption{Mean precision, recall and NDCG for different values of $k$ for the FM recommender, the popularity baseline and the MF-like baseline. \label{fig:results}}
\end{figure}

Figure~\ref{fig:results} shows the results. As we would expect, precision falls but recall and NDCG rise as $k$ increases. But what is important is the comparison with the baselines: FM outperforms them for
every metric, at every recommendation list size, $k$.
As an example, given that FM's precision at $k=5$ is nearly 60\%, 3 of the 5 suppliers to a new new event will, on average, be relevant to the event. By contrast, the popularity recommender's precision of about 35\% means that only 1 or 2 of its recommendations will be relevant on average.

These FM results are impressive given the limitations of the meta-data that we are using at the moment. For example, we do not have any features that explicitly describe a supplier's geographical coverage, which might be strongly predictive of relevance. Instead, FM is doing well at predicting relevance using text descriptions (which, by manual inspection, we note are often not very informative) and from the learned interactions between features and historical participation data.

\section{Discussion} \label{sec:disc}

The results of the experiment are promising. They certainly encourage us to consider more systematic recording of event participation data and event meta-data to give a dataset large enough to support more detailed experiments (e.g.\ the comparison of more recommendation algorithms) in the future.

\begin{figure}[t] 
\includegraphics[width=0.45\textwidth]{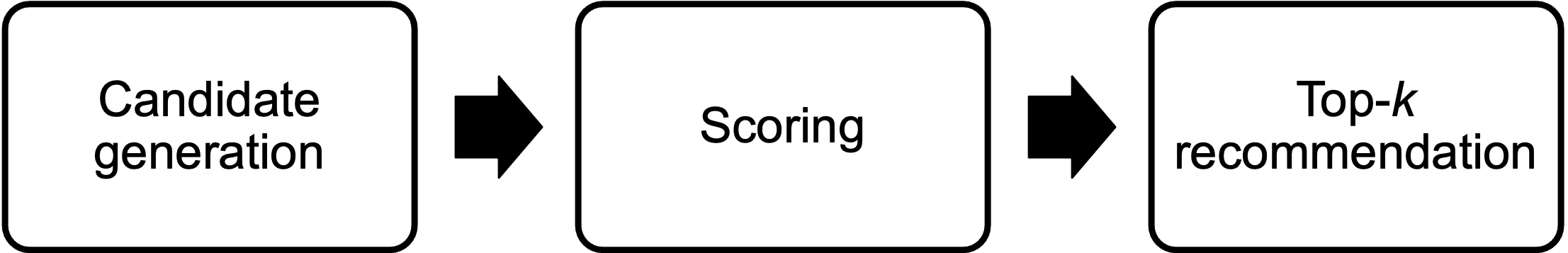}
\caption{Recommendation in 3 Steps\label{fig:steps}}
\end{figure}

But there are some broader issues that have arisen during the work that we think it useful to discuss here. We refer to Figure \ref{fig:steps}, which presents recommendation as a 3-step process. We acknowledge this is highly simplified, but we are using it only to structure the discussion; it is not intended as a system architecture.

The paper has focused on the second step, Scoring: given a new event and a set of candidate suppliers, the recommendation algorithm predicts the relevance of the candidates to the event. But the first and third steps deserve discussion.

The first step generates the set of candidates. It is na\"{\i}ve to assume, as we have done in the experiments in this paper, that the candidates should be \emph{all} suppliers known to the system. More realistically, filters can be applied: there may be geographical constraints (e.g.\ only include suppliers on the same landmass as the event); there may be service-level constraints (e.g.\ only include suppliers who can offer refrigerated delivery); there may be business constraints (e.g.\ to include or exclude suppliers for commercial reasons). The job of the recommendation algorithm, i.e.\ scoring the candidates, may be made simpler if the set of candidates has already been filtered in the first step. A serious consideration of this step is needed if the ideas are to be moved towards deployment.

The third step selects the top-$k$ items for recommending to the user (the purchaser who has created this event). In the experiments in this paper, we assumed that we simply select the $k$ candidates that received the highest relevance scores by the recommendation algorithm in the second step. But this is also na\"{\i}ve. It may be important to balance relevance with other criteria, especially criteria that consider the top-$k$ as a whole. For example, we may want to ensure a degree of diversity among the members of the top-$k$. Methods for doing this in recommender systems are surveyed in, e.g., \cite{Kaminskas-Bridge-2017}. Another example criterion that has been explored in recommender systems research more recently is fairness; see, e.g., \cite{Abdollahpouri-Et-Al-2020} for a survey of this work. Finally, there may again be business rules that affect the make-up of the top-$k$. This step also needs serious consideration prior to deployment of the ideas in this paper.

\section{Conclusions} \label{sec:concs}

In this work, we proposed a recommender system 
to assist supplier discovery in online procurement. This is a largely novel domain, with limited data at the moment.
Our approach uses event meta-data and historical event participation data, producing more accurate
recommendations for cold-start events (which are the main use case in this domain) when compared to a non-personalized
popularity baseline and an approach that is closer to standard matrix factorization.

While already offering a satisfactory performance,
the model could be improved by the addition of
more event meta-data, such as descriptions of the supplier and purchaser companies, for instance.
It would also benefit from more historical event participation data. 
With more data, it would then make sense to explore a wider range of recommendation algorithms, including ones that balance recommendation relevance with top-$k$ diversity or fairness. 

More data would also enable us to model the problem at a finer level of granularity. At the moment, we are recommending suppliers to events. But, in reality, in an event a purchaser may have several different requirements, sometimes called lots (or even lanes). It does not follow that the same supplier should be used for all the lots in an event. With more data, we could compare our current approach (recommending suppliers to events) to a finer-grained formulation (recommending suppliers to lots).


\bibliographystyle{IEEEtran}
\bibliography{main}

\end{document}